\documentclass[review,a4paper,10pt,onecolumn]{elsarticle}

\usepackage{lineno,hyperref}
\usepackage{geometry}
\usepackage{float}

\begin{document}

\begin{frontmatter}

\title{Characterizing the temporal rotation and radial twist of the interference pattern of vortex beam}

\author[mymainaddress,mythirdaddress]{Longzhi Nie}

\author[mymainaddress,mythirdaddress]{Lingran Kong}

\author[mymainaddress]{Tianyou Gao \corref{mycorrespondingauthor}}
\cortext[mycorrespondingauthor]{Corresponding author}
\ead{602gty@sina.com}

\author[mymainaddress,mythirdaddress]{Nenghao Dong}

\author[mymainaddress,mysecondaryaddress]{Kaijun Jiang \corref{mycorrespondingauthor}}
\cortext[mycorrespondingauthor]{Corresponding author}
\ead{kjjiang@wipm.ac.cn}

\address[mymainaddress]{State Key Laboratory of Magnetic Resonance and Atomic and Molecular Physics, Wuhan Institute of Physics and Mathematics, Innovation Academy for Precision Measurement Science and Technology, Chinese Academy of Sciences, Wuhan 430071, China}
\address[mysecondaryaddress]{Center for Cold Atom Physics, Chinese Academy of Sciences, Wuhan 430071, China}
\address[mythirdaddress]{University of Chinese Academy of Sciences, Beijing 100049, China}

\begin{abstract}
 We experimentally and theoretically characterize the temporal rotation and radial twist of the interference pattern of vortex beam with its conjugate copy. To quantitatively study the temporal rotation and radial twist, we controllably modify the conjugate beam with a frequency or wavefront curvature difference using a movable Mach-Zehnder interferometer. The effects of the physical parameters (i.e., the topological charge, frequency difference and wavefront curvature difference of the vortex beams) on the temporal rotation as well as radial twist are systematically explored. We further measure two parameters, the rotation velocity $\Omega$ and twist coefficient $\alpha$, respectively, to characterize the degree of the temporal rotation and radial twist of the interference pattern. The theory of the interference model on vortex beams have good agreements with the experimental results. This work is favorite to study the detailed structure of the interference pattern and manipulate matter with superimposed vortex beams.
\end{abstract}


\end{frontmatter}

\section{Introduction}
Vortex laser beams possessing an azimuthal phase structure $\exp(il\phi)$ carry orbital angular momentum (OAM) of $l\hbar$ per photon, where $l$ is an integer number, referred to as
the topological charge (TC) \cite{Allen1992PRAoam, Padgett1995OC}. Laguerre-Gaussian (LG) beam is a typical vortex optical beam. The magnitude of TC expresses the number of $2\pi$ phase cycles around the singularity, and the sign indicates the rotation direction of the wavefront around the central axis. Due to its special property with controllable OAM, the vortex optical beam has been extensively used in diverse research areas, including optical communication \cite{Bozinovic2013Science, Vallone2014prl}, optical imaging \cite{Yao2011Adv}, manipulation of mesoscopic objects \cite{Grier2003Nature, Paterson2001Science, Lehmuskero2014OE}, and production of vortex states in ultracold quantum gases \cite{Pu2015PRAsoam, Zhang2015PRAsoam, Lin2018PRLsoam, Lin2018PRLrotation, Jiang2019PRLSOAMC}.

Interferometry is a fundamental method to explore the characters of the vortex laser beam and its applications. The structure of the interference pattern strongly depends on relative parameters of vortex beams. When there exists a frequency or wavefront curvature difference between two interfering vortex beams, the temporal rotation or radial twist will emerge in the interference pattern \cite{Leach2007OE, Arlt2002OEMovingDoppler, Harris1994OCspiral}, respectively, which has potential applications. For example, the twisted interference fringes were proposed to enhance the gradient force exerted by the Tornado-like pattern upon micro-particles in optical tweezers \cite{Hu2021PhysicsVortices} and measure the curvature radius of a concave mirror or convex lens \cite{Liu2021OpticsRadius}. The rotating interference pattern has broad applications in manipulation of cold atoms, such as generating artificial gauge electromagnetic field \cite{Aldossary2017JOSAArtificial}, exploring quantum Hall physics \cite{Peter2016PRAHall}, rotating single atoms \cite{Zhan2009OEZhan}, and so on. Previous works have observed the rotation and twist effects of the interference pattern mainly based on the visual inspection \cite{Leach2007OE, Arlt2002OEMovingDoppler, Harris1994OCspiral, Cui2019OE, Vickers2008OSA}. The knowledge of the detailed structure of the optical interference pattern is required to understand the interaction between matter and superimposed vortex beams. Quantitatively characterizing the temporal rotation and radial twist of the interference pattern is favorite to extend the applications of vortex beam.

In this paper, we quantitatively characterize the temporal rotation and radial twist of the interference pattern of vortex beam with its conjugate copy. To study the radial twist, we modify the conjugate beam with a wavefront curvature difference using a movable Mach-Zehnder (MZ) interferometer. To study the temporal rotation, we modify the conjugate beam with a frequency difference using two acousto-optic modulators (AOMs). The effects of the physical parameters (i.e., the TC, frequency difference and wavefront curvature difference of the vortex beams) on the temporal rotation as well as the radial twist are explored. We further measure two parameters, the rotation velocity $\Omega$ and twist coefficient $\alpha$, respectively, to characterize the degree of the temporal rotation and radial twist of the interference pattern. The theory of the interference model on vortex beams have good agreements with the experimental results. Here we study the interference between the vortex beam and its conjugate copy. The method demonstrated is also applicable to other kinds of vortex beams.

\section{Theoretical analysis} \label{Sec.1}
The vortex beam contains a phase singularity where the intensity vanishes \cite{Khajavi2017OL, Riley1977ApplOpt}. For a LG beam, the scaled electric field is
  \begin{equation}\label{Eq.(1)}
  E(r,\phi,z,t)=A(r)\exp[-i\theta(r,z)]\exp(-i\omega t)\exp(-il\phi).
  \end{equation}

  \begin{equation}\label{Eq.(2)}
  \theta(r,z)=\frac{k}{2}\frac{r^2}{R(z)}+(2p+|l|+1)\arctan(\frac{z}{z_R})+kz+\vartheta,
  \end{equation}
  \begin{equation}\label{Eq.(3)}
  A(r)=E_0C_{lp}^{LG}L_p^{|l|}(\frac{2r^2}{w^2(z)})\frac{w_0}{w(z)}(\frac{r\sqrt{2}}{w(z)})^{|l|}\exp(-\frac{r^2}{w^2}),
  \end{equation}
\noindent where $\omega$ is the laser frequency, $r$ is the radius, $\phi$ is the azimuthal angle, $z$ is the propagation distance, $w(z)=w_0\sqrt{1+(z/z_R)^2}$ is the beam radius with the waist radius $w_0$, $z_R=\pi w_0^2/\lambda$ is the Rayleigh length with the wavelength $\lambda$, $k=2\pi/\lambda$ is the wave vector, $L_p^{|l|}$ is an associated Laguerre polynomial, $C_{lp}^{LG}=\sqrt{\frac{2p!}{\pi(p+|l|)!}}$is the normalization coefficient, $\vartheta$ represents the additional phase shift induced by the optics in the optical path, $p$ is the radial index, and $l$ is the azimuthal index (referred as the TC) \cite{Chappuis2017pra}.
The radius of the wavefront curvature is
  \begin{equation}\label{Eq.(4)}
  R(z)=z(1+\frac{z_R^2}{z^2}).
  \end{equation}

We consider the simple case with $p=0$. By interfering two co-propagating LG beams ($i=1,\ 2$), we obtain the intensity distribution \cite{Leach2007OE, Lembessis2017pra},
  \begin{equation}\label{Eq.(5)}
  I=\frac{n c\varepsilon}{2}|E_{1}+E_{2}|^2 = \frac{n c\varepsilon}{2} \left[A_{1}(r)^2+A_{2}(r)^2+2A_{1}(r)A_{2}(r)\cos (-\Delta l\phi-\Delta\omega t-\Delta\theta)\right],
  \end{equation}
where $\Delta l = l_1 - l_2$, $\Delta\omega=\omega_{1}-\omega_{2}$ and $\Delta\theta=\theta_{1}-\theta_{2}$. $\varepsilon$ is the dielectric constant, $n$ is the refractive index, and $c$ is the speed of light. The azimuthal angle for the maximum intensity of the bright interference fringe is
  \begin{equation}\label{Eq.(6)}
      \phi_{max}(t)=-\frac{2m\pi + \Delta\omega t+\Delta\theta}{\Delta l},
  \end{equation}
\noindent where $m$ is an integer.

Here we consider that a vortex beam interferes with its conjugate copy. Then $l_2= -l_1$ and $\Delta l = 2l_1$. From the first term on the right of Eq. \ref{Eq.(6)}, the number of the bright interference fringes in one circle equals $|\Delta l|$, twice the magnitude of the TC. The temporal rotation and radial distribution of the bright interference fringes depend on $\Delta\omega$, $\Delta\theta$ and $\Delta l$, which are related to the other two terms of Eq. \ref{Eq.(6)}. So we will analyze the structure of the interference pattern from two aspects as follows.

First, when the conjugate light is modified with a frequency difference $\Delta\omega \neq 0$, $\phi_{max}$ is time-dependent, and the whole interference pattern will rotate around the central axis. The rotation velocity is
  \begin{equation}\label{Eq.(7)}
  \Omega=\frac{\partial\phi_{max}}{\partial t}=-\frac{\Delta\omega}{\Delta l}.
  \end{equation}
\noindent Adopting the convention of the right-hand coordinate system, when signs of $\Delta\omega$ and $\Delta l$ are the same, the interference pattern will rotate clockwise ($\Omega < 0$), and vice versa. So with the knowledge of $\Delta\omega$ and $\Delta l$, we can determine the temporal rotation of the interference pattern.

Secondly, $\Delta\theta$ in Eq. \ref{Eq.(6)} can be written as
  \begin{equation}\label{Eq.(8)}
  \Delta\theta=\Phi(z_{1},z_{2})+\frac{k}{2}(\frac{1}{R_{1}(z)}-\frac{1}{R_{2}(z)})r^2,
  \end{equation}
\noindent where the term $\Phi$ varying with $z$ is the so-called Gouy phase, and its influence on interference pattern has been discussed in reference \cite{Baumann2009OE}. $\phi_{max}$ is $r$-dependent. In the case $\Delta\omega = 0$, $\phi_{max}$ is written as
  \begin{equation}\label{Eq.(9)}
  \phi_{max}=\alpha r^2-\frac{2m\pi+\Phi(z_{1},z_{2})}{\Delta l},
  \end{equation}
  \begin{equation}\label{Eq.(10)}
  \alpha=-\frac{k}{2}\frac{\Delta\rho}{\Delta l}=-\frac{k}{2}\frac{\Delta(\frac{1}{R})}{\Delta l}=-\frac{k}{2} \frac{\frac{1}{R_{1}(z)}-\frac{1}{R_{2}(z)}}{\Delta l},
  \end{equation}
\noindent where $\rho=\frac{1}{R}$ is the wavefront curvature. When the conjugate light is modified with a wavefront curvature difference $\Delta\rho \neq 0$, the interference fringe twists along the radial direction. $\alpha$ is defined as the twist coefficient. If the signs of $\Delta\rho$ and $\Delta l$ are the same, i.e., $\alpha<0$, the interference fringe twists clockwise, and vice versa. So with the knowledge of $\Delta\rho$ and $\Delta l$, we can determine the radial twist of the interference fringe. The magnitude of $\alpha$ characterizes the degree of the radial twist.

\section{Experimental setup}
\begin{figure}[htbp]
\centerline{\includegraphics[width=14cm]{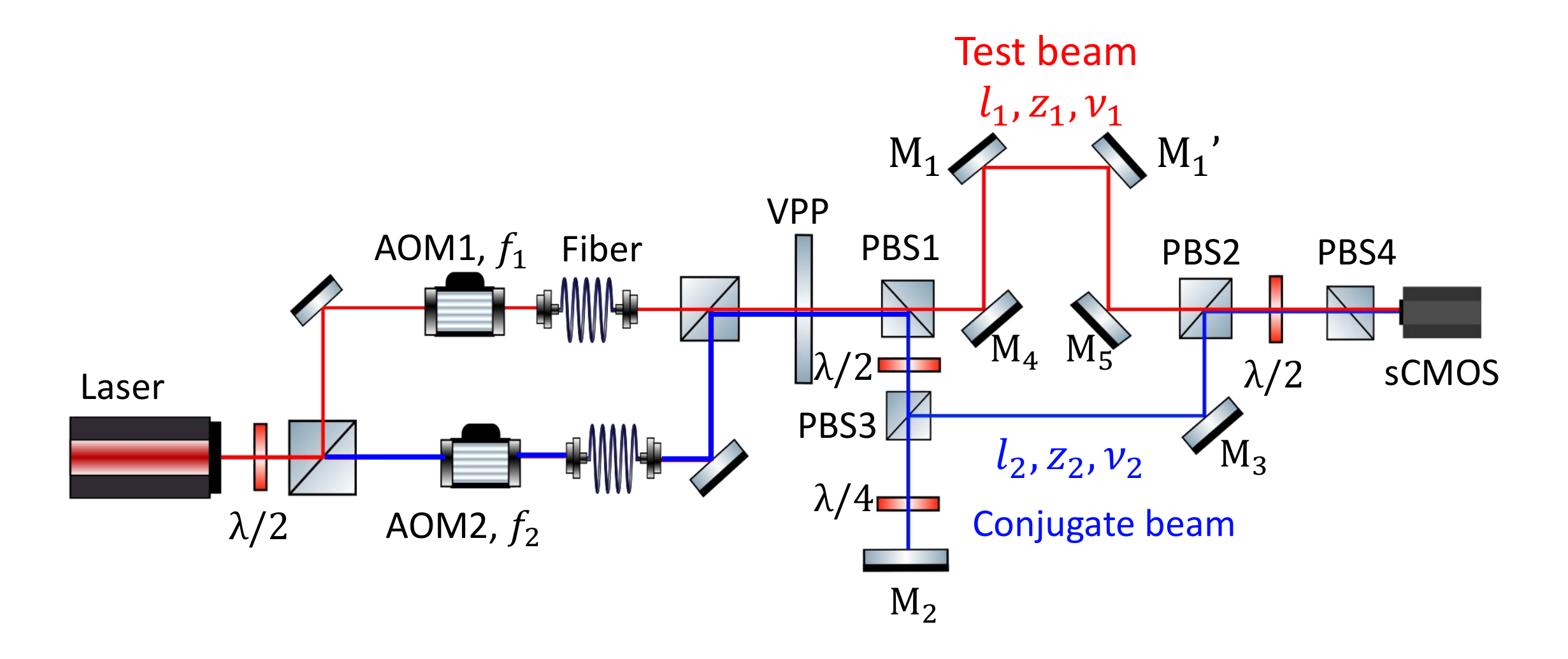}}
\caption{Experimental setup to measure the interference pattern of the vortex beam with its conjugate copy. The vortex beams are generated after the VPP. The test vortex beam (red lines) and its conjugate copy (blue lines) form a MZ interferometer. $l_{1}$ and $l_{2}$ are the TCs of the two beams, respectively. The interference pattern is probed with a sCMOS camera. Two AOMs (AOM1 with the driving frequency $f_{1}$, AOM2 with $f_{2}$) control the frequencies ($\nu_{1}$, $\nu_{2}$) of the two beams, respectively. The positions of the mirrors M$_{1}$ (M$_{1}^{'}$) and M$_{2}$ control the propagation distances ($z_{1}$, $z_{2}$) of the two beams, respectively. VPP, vortex phase plate. AOM, acousto-optic modulator. PBS, polarization beam splitter. $\lambda/2$, half-wave plate. $\lambda/4$, quarter-wave plate.}  \label{fig.1}
\end{figure}

We has theoretically analyzed the temporal rotation and radial twist of the interference pattern in Section \ref{Sec.1}. Then we will observe these two behaviors in experiment. Fig. \ref{fig.1} schematically presents the experimental setup. The laser with the wavelength $\lambda=782.7$ nm is coupled through the fiber to obtain a Gaussian beam. A vortex phase plate (VPP) (RPC Photonics, VPP-m780) is used to convert a Gaussian beam into a LG beam with a TC of $l_{1}$. The main setup is a MZ interferometer. The sign of the TC will change once the LG beam is reflected. Then the transmitted beam from polarizing beam splitter 1 (PBS1) is reflected four times (M$_{4}$, M$_{1}$, M$_{1}^{'}$, M$_{5}$), and the TC is still $l_{1}$ after PBS2. We call this beam the test beam. The reflected beam from PBS1 is reflected five times (PBS1, M$_{2}$, PBS3, M$_{3}$, PBS2), and becomes the conjugate beam after PBS2, i.e., $l_{2}=-l_{1}$. The method that we change the sign of the TC by controlling the times of the reflection is simpler than using a Dove prism \cite{Vickers2008OSA, Nie2015OCmeasureTC}. Also it is convenient to adjust the wavefront curvature difference between the test and conjugate beams, which will be explained later. We use two AOMs (AOM1 with the driving frequency $f_{1}$, AOM2 with $f_{2}$) to control the frequencies of the two beams, respectively. The frequency of the test beam is $\nu_{1}=f_{1}$, and $\nu_{2}=f_{2}$ for the conjugate beam. Then $f_{1}-f_{2}=\nu_{1}-\nu_{2}=\Delta\nu$. The polarizations of the two beams after PBS4 are the same, and the interference pattern between the two beams is probed with a sCMOS camera (pco.edge 4.2).

According to Eq. \ref{Eq.(4)}, $R(z)$ is determined by the propagation distance $z$. VPP is placed at the focus of the beam ($z=0$). The waist radius is $w_0=4.2\times10^{-4}$ m and the Rayleigh length is $z_R=0.7$ m. We set $z>z_{R}$ in the experiment, then there is a one-to-one correspondence between $R(z)$ and $z$. The mirrors (M$_{1}$, M$_{1}^{'}$ and M$_{2}$) are movable. We control the distances ($z_{1}$, $z_{2}$) of the test beam and its conjugate copy by adjusting the positions of mirrors M$_{1}$ (M$_{1}^{'}$) and M$_{2}$, respectively. $\Delta z = z_{1}-z_{2}$. The wavefront curvature difference between the two beams can be written as
\begin{equation}\label{Eq.(11)}
\Delta\rho=\frac{1}{R_{1}(z)}-\frac{1}{R_{2}(z)}=\frac{-\Delta z({z_{1}z_{2}}-z_{R}^{2})}{(z_{1}^{2}+z_{R}^{2})(z_{2}^{2}+z_{R}^{2})}.
\end{equation}
\noindent Under the condition $z_{1, 2} > z_{R}$, the signs of $\Delta z$ and $\Delta\rho$ are opposite. We control the value of $\Delta\rho$ by adjusting $\Delta z$.

\section{Results and discussion}
\subsection{Temporal rotation of the interference pattern}
\begin{figure}[htbp]
\centerline{\includegraphics[width=14cm]{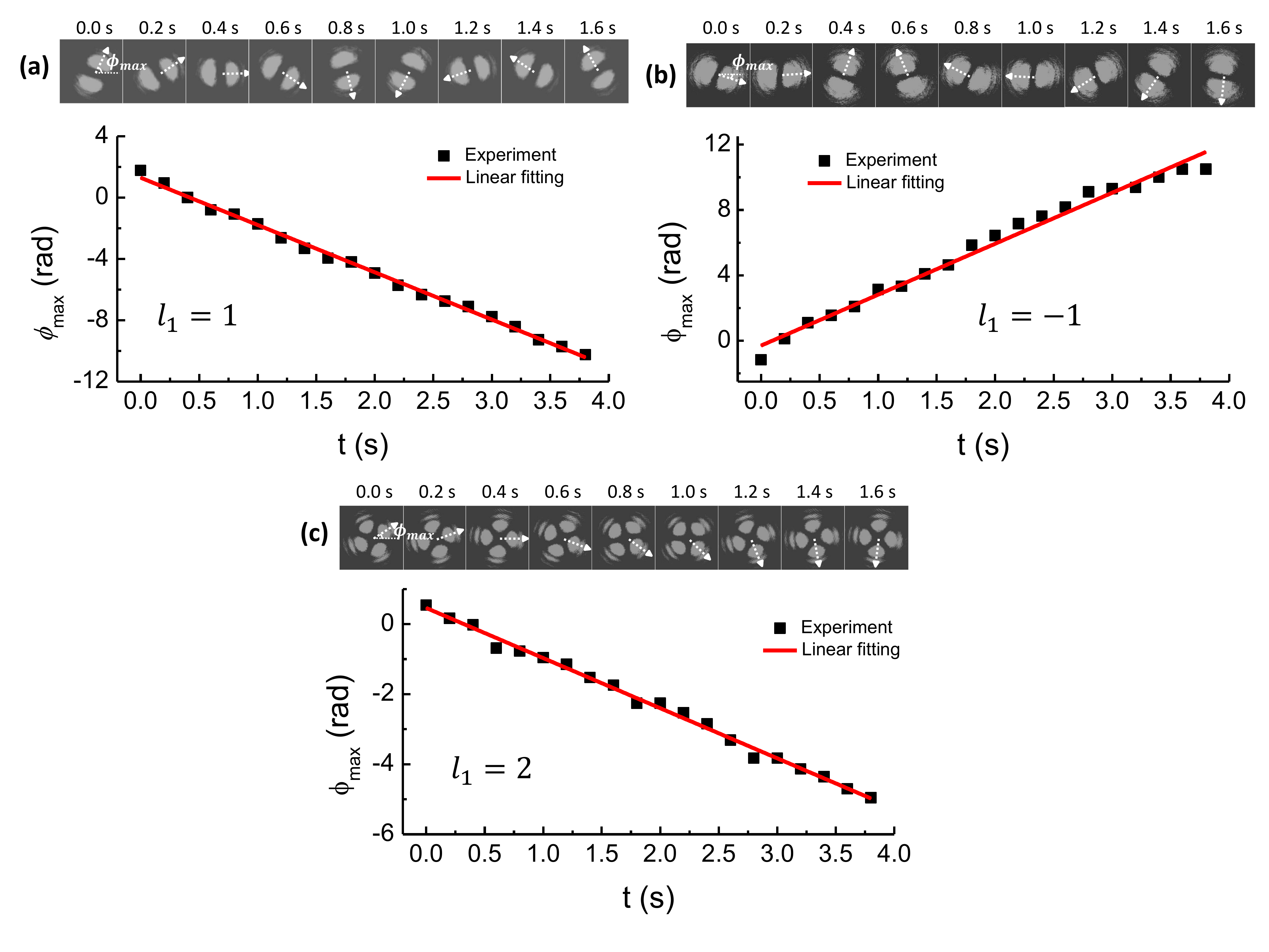}}
\caption{Temporal rotation of the interference pattern with a frequency difference $\Delta\nu=\nu_{1}-\nu_{2}=1$ Hz. (a) is for $l_1=1$, (b) for $l_1=-1$ and (c) for $l_1=2$. The upper panel shows the rotating patterns imaged every 0.2 s. $\phi_{max}$ is the azimuthal angle for the maximum intensity of the bright interference fringe. The lower panel shows $\phi_{max}$ as a function of the rotating time. The red solid line is a linear fitting used to extract the rotation velocity $\Omega$. From the fitting, $\Omega= \left(-3.11(9),\ 3.04(3),\ -1.61(3) \right)$ rad/s for $l_1= \left(1,\ -1,\ 2 \right)$. }\label{fig.2}
\end{figure}

Fig. \ref{fig.2} shows the rotation of the interference pattern with a frequency difference $\Delta\nu=1$ Hz. Here $z_{1}=z_{2}=1.4$ m, and the interference fringe has no radial twist. The number of interference fringes equals $2|l_{1}|$. The interference pattern rotates clockwise with $l_{1} > 0$ (i.e., $l_{1} = 1, \ 2$). When the sign of $l_1$ changes (i.e., $l_{1} = -1$), the rotation direction is reversed. Moreover, the rotation for $l_{1} = 1$ is faster than that for $l_{1} = 2$. All these observations are consistent with predictions of Eq. \ref{Eq.(7)}. $\phi_{max}$ is the azimuthal angle for the maximum intensity of the bright interference fringe. The detail to determine $\phi_{max}$ is shown in Fig. \ref{fig.6}. To extract the value of rotation velocity $\Omega$, we measure the time evolution of $\phi_{max}$, as shown in the lower panel of Fig. \ref{fig.2}. We use a linear function $\phi_{max}=\Omega t+\phi_0$ to fit the experimental data, where $\Omega$ and $\phi_0$ are the fitting parameters. From the fitting, $\Omega= -3.11(9)$ rad/s for $l_1= 1$, $\Omega= 3.04(3)$ rad/s for $l_1= -1$, and $\Omega= -1.61(3)$ rad/s for $l_1= 2$. The value in the parenthesis is the standard deviation from the fitting. From the calculation of Eq. \ref{Eq.(7)}, $\Omega= \left(-3.14,\ 3.14,\ -1.57 \right)$ rad/s for $l_1= \left(1,\ -1,\ 2 \right)$. The theoretical calculations have good agreements with the experimental measurements.

\begin{figure}[htbp]
\centerline{\includegraphics[width=14cm]{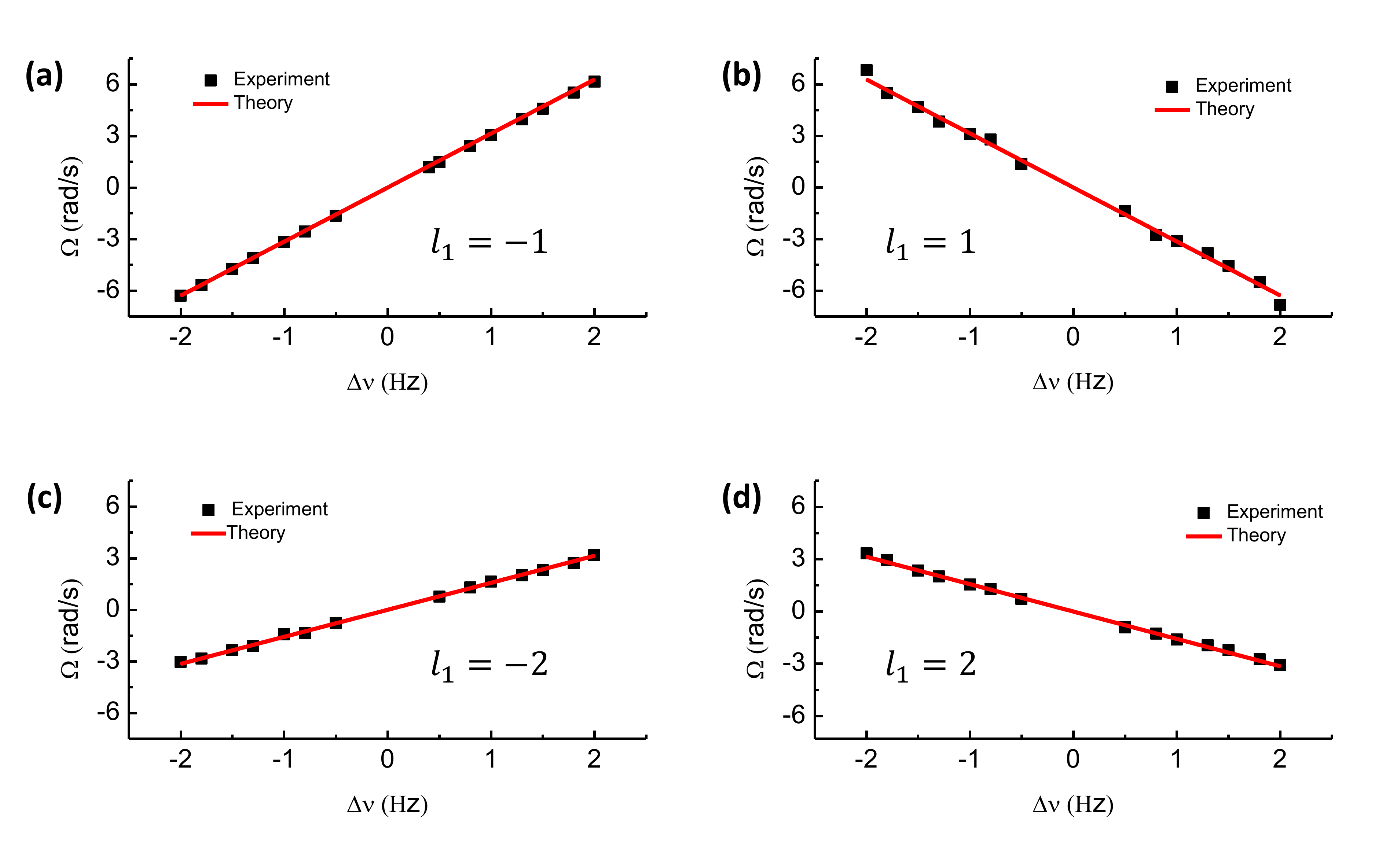}}
\caption{Rotation velocity $\Omega$ for positive and negative TCs. $\Omega$ is plotted versus $\Delta\nu$. (a) is for $l_1=-1$, (b) for $l_1=1$, (c) for $l_1=-2$, and (d) for $l_1=2$. The error bars are smaller than the data marks. The red solid line is the theoretical calculation of Eq. \ref{Eq.(7)}.}\label{fig.3}
\end{figure}

In Fig. \ref{fig.3}, we measure $\Omega$ for positive and negative TCs (i.e., $l_1= \pm1,\ \pm2$). $\Delta\nu$ varies from $-2$ Hz to $2$ Hz. When $\Delta\nu$ is small ($\Delta\nu\approx0$), the rotation is too slow to be probed. It is shown that when the signs of $l_{1}$ and $\Delta\nu$ are the same, $\Omega < 0$, and vice versa. $\Omega$ increases (decreases) linearly versus the increasing of $\Delta\nu$ when $l_{1}$ is negative (positive). The calculation of Eq. \ref{Eq.(7)} agrees well with the experimental results.

Previous works have observed the images of the rotating interference pattern with a frequency difference \cite{Leach2007OE, Arlt2002OEMovingDoppler}, while the quantitative dependence of the rotation on the physical parameters is yet to be explored. Here we measure the temporal evolution of the parameter $\phi_{max}$. We further measure the rotation velocity $\Omega$ to quantitatively characterize the temporal rotation of the interference pattern versus the frequency difference $\Delta\nu$ as well as the TC $l_1$.

\subsection{Radial twist of the interference fringe}

\begin{figure}
\centerline{\includegraphics[width=10cm]{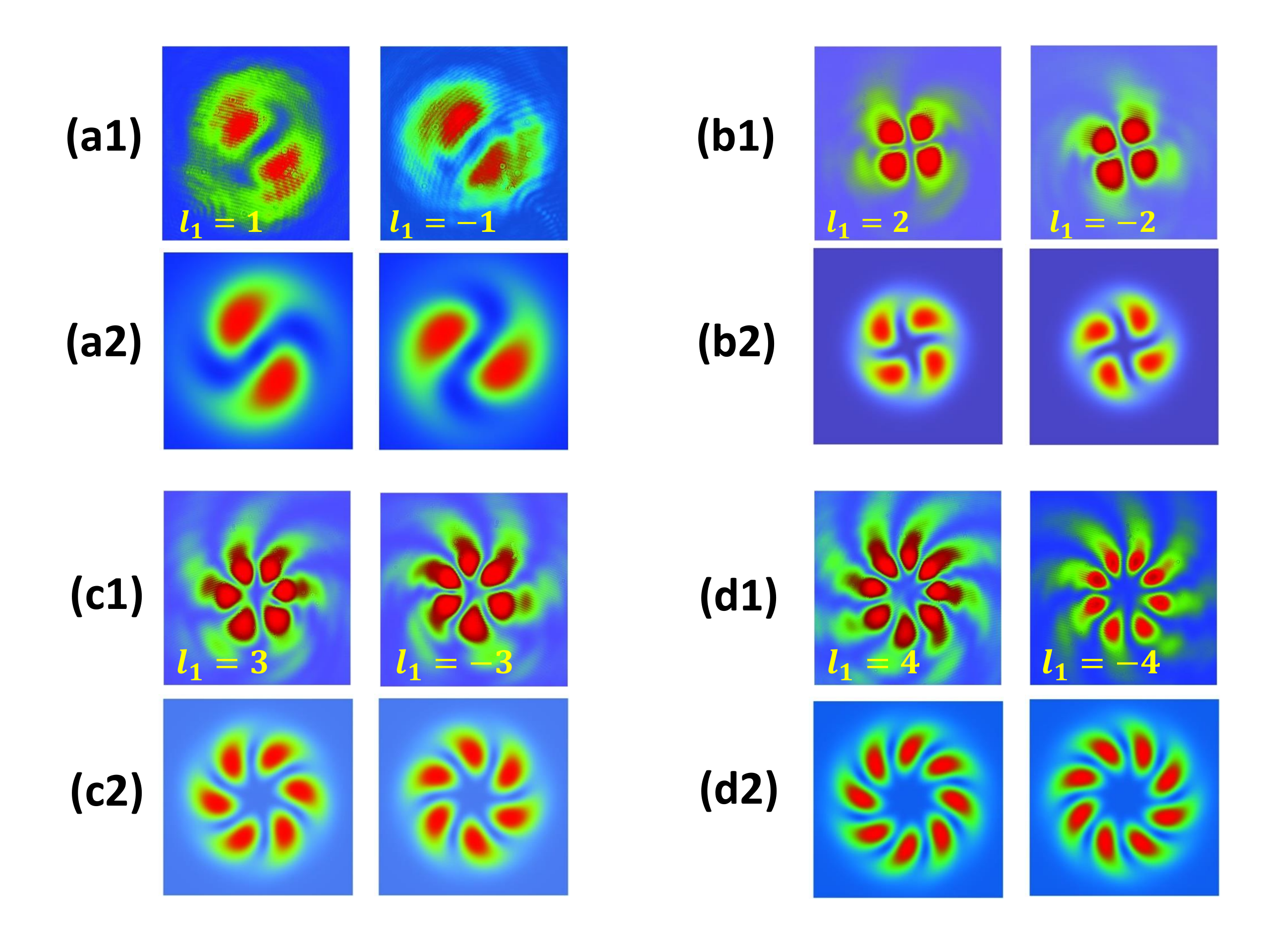}}
\caption{Twisted interference patterns with a wavefront curvature difference. (a1), (b1), (c1) and (d1) are the experimental observations for $l_1=\pm1, \pm2, \pm3, \pm4$, respectively. (a2), (b2), (c2) and (d2) are the corresponding theoretical simulations with Eq. \ref{Eq.(5)}. In (a1), $z_{1}=1.1$ m and $z_{2}=2.25$ m. In (b1), (c1) and (d1), $z_{1}=1.4$ m and $z_{2}=2.0$ m.}\label{fig.4}
\end{figure}

To study the radial twist of the interference fringe, we set $\Delta\nu = 0$. In this case, the interference pattern is static. Fig. \ref{fig.4} shows the twisted interference patterns with a wavefront curvature difference $\Delta \rho \neq 0$. Under the condition $\Delta z < 0$ (i.e., $\Delta \rho > 0$), if $l_1$ is positive (i.e., $l_1 = 1, \ 2,\ 3,\ 4$ ), the interference fringe twists clockwise ($\alpha <0$). When the sign of $l_1$ changes (i.e., $l_1 = -1, \ -2,\ -3,\ -4$ ), the the interference fringe twists anti-clockwise ($\alpha >0$). These twist effects can also be predicted by Eq. \ref{Eq.(9)} and \ref{Eq.(10)}. The calculated interference patterns with Eq. \ref{Eq.(5)} also show obvious radial twists, similar to the experimental results. It is noted that for $l_1=\pm1$, $\Delta z$ is big to clearly show the twist tails.

\begin{figure}
\centerline{\includegraphics[width=10cm]{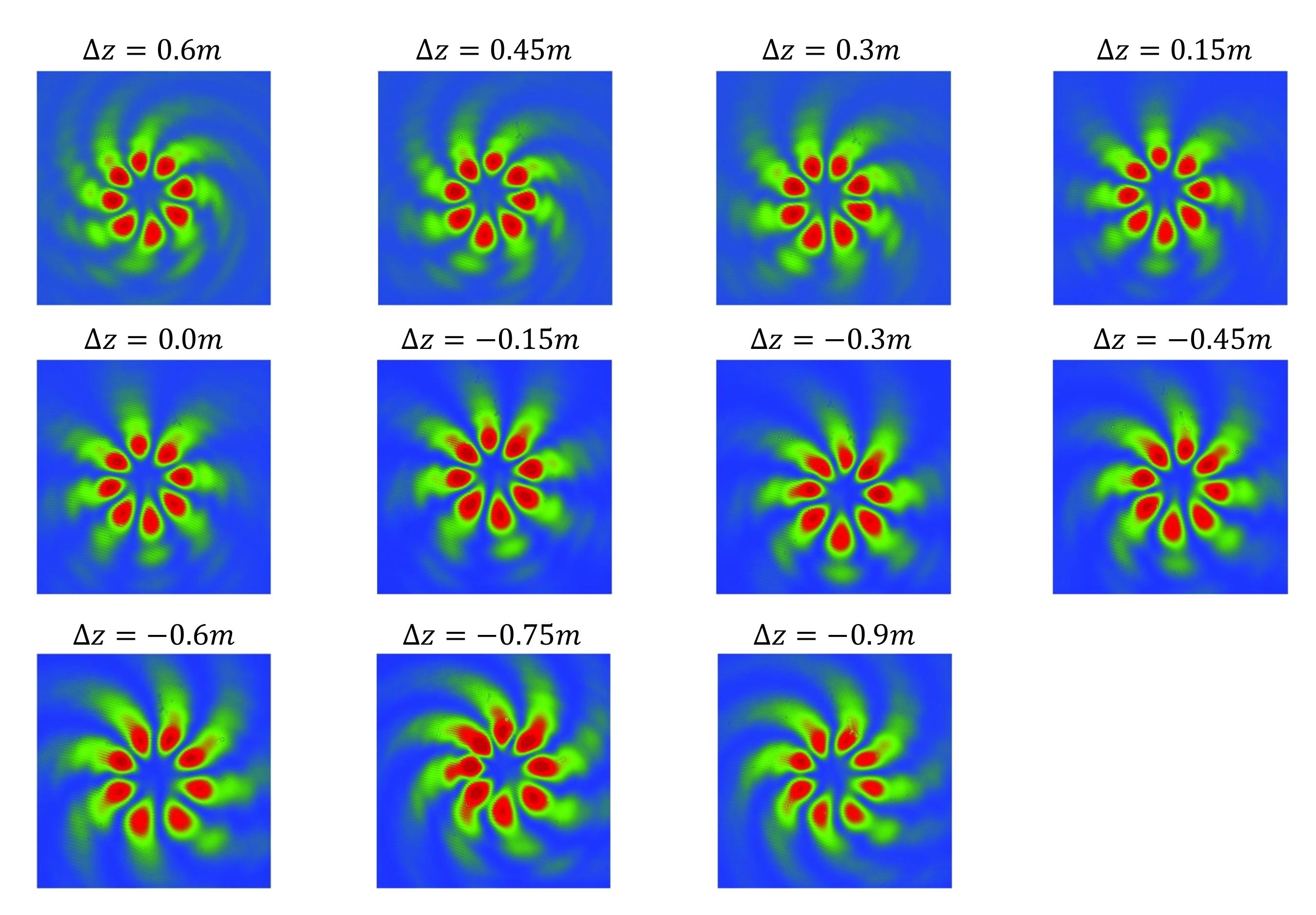}}
\caption{Twisted interference patterns with different values of $\Delta z$. $\Delta z$ decreases from 0.6 m to -0.9 m with a step 0.15 m. $l_1=-4$.}\label{fig.5}
\end{figure}

Fig. \ref{fig.5} schematically shows the effect of the wavefront curvature difference on the radial twist of interference fringes. $z_{1}=1.4$ m is fixed, and we change $z_{2}$ to get different values of $\Delta z$. The twist direction of interference fringes is reversed from the clockwise to anti-clockwise when $\Delta z$ varies from the positive to negative. The twist effect becomes more obvious with increasing the magnitude of $\Delta z$. For $\Delta z = 0$, the twist effect vanishes. These observations are consistent to the predictions of Eq. \ref{Eq.(9)} and \ref{Eq.(10)}.

\begin{figure}[htbp]
\centerline{\includegraphics[width=12cm]{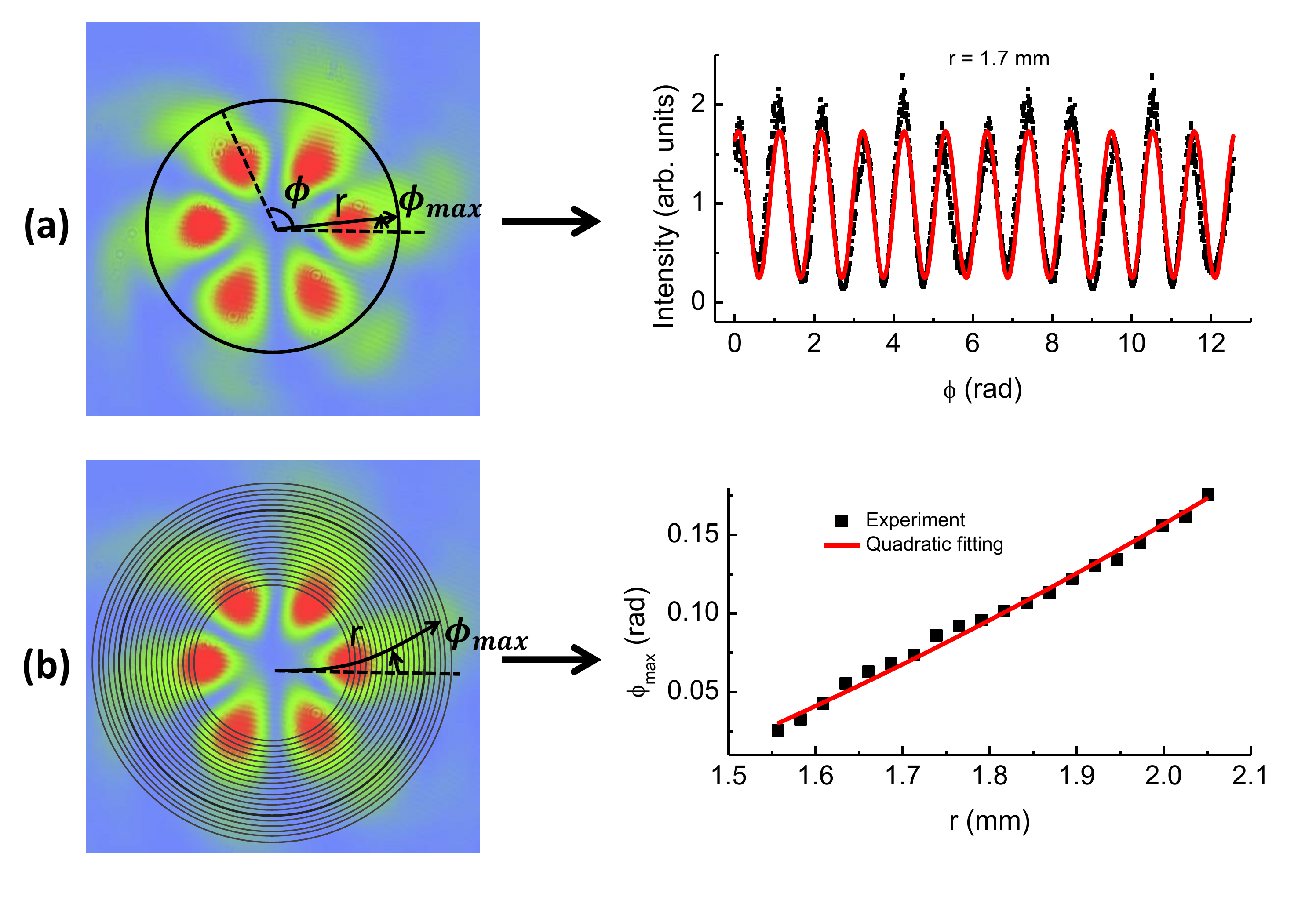}}
\caption{Measuring the twist coefficient $\alpha$. $l_{1}=-3,\ \Delta z=-0.6$ m. (a) denotes the angular interference fringe with the radius $r=1.7$ mm. $\phi_{max}$ is the azimuthal angle for the maximum intensity of the first bright interference fringe. We use a cosine function (red solid curve) to numerically fit the experimental data, extracting $\phi_{max}=0.086$ rad. (b) plots $\phi_{max}$ versus $r$. The red solid curve is the quadratic fitting of Eq. \ref{Eq.(9)}, which gives $\alpha=8.05(17)\times10^4$ rad m$^{-2}$. The theoretical value of Eq. \ref{Eq.(10)} is $\alpha=8.30\times10^4$ rad m$^{-2}$. The black solid circles one the left column schematically indicate the interference paths.}\label{fig.6}
\end{figure}

\begin{figure}[htbp]
\centerline{\includegraphics[width=10cm]{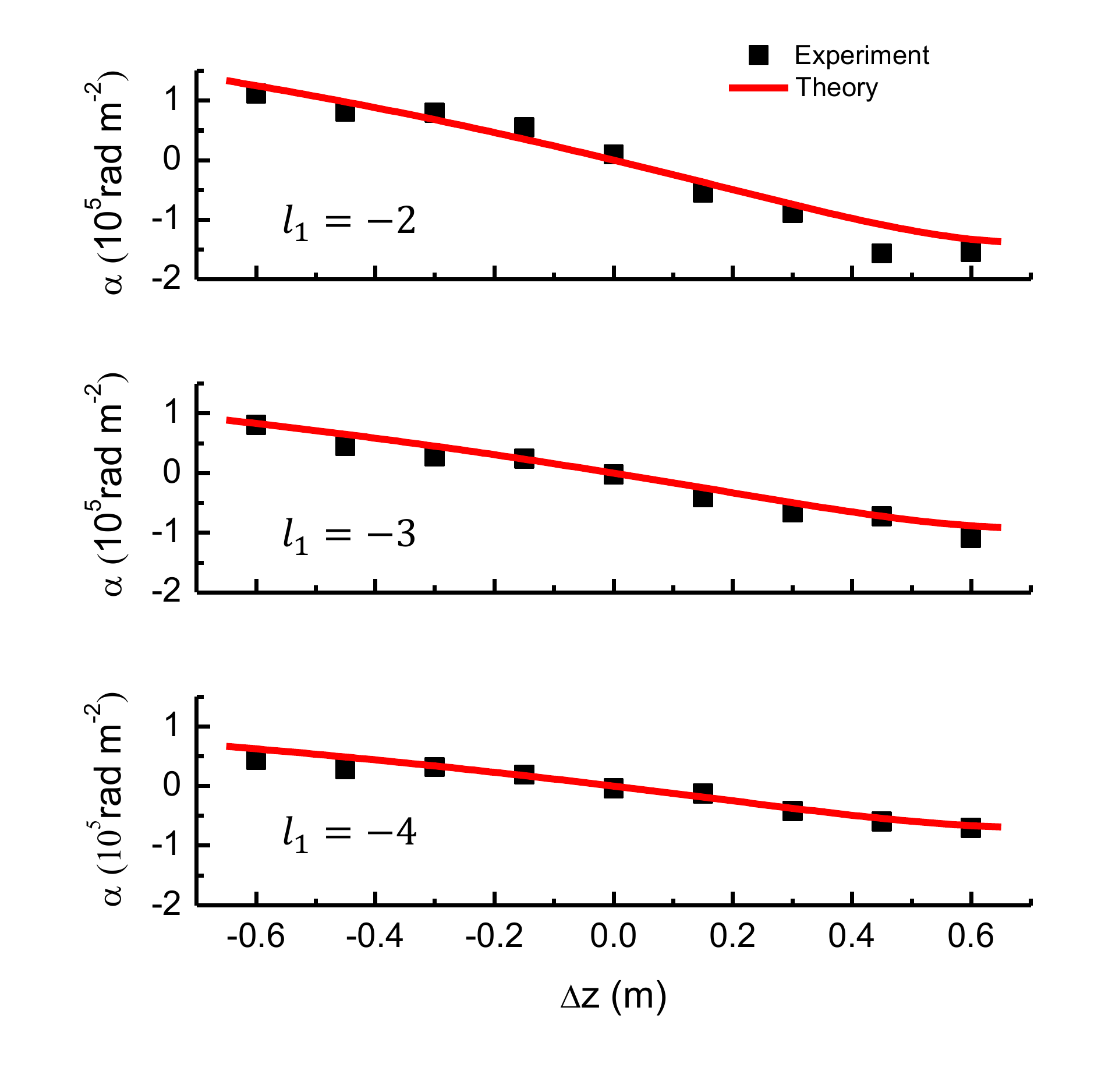}}
\caption{Twist coefficient $\alpha$ for different TCs. $l_{1}=-2,\ -3,\ -4$. $\alpha$ is measured as a function of $\Delta z$. The error bars are smaller than the data marks. The red solid curves are the theoretical calculations of Eq. \ref{Eq.(10)}.}\label{fig.7}
\end{figure}

To quantitatively characterize the twist degree of the interference fringe, we measure the twist coefficient $\alpha$ which quantifies the variation of $\phi_{max}$ along the radial direction. We take the interference pattern with $l_{1}=-3$ as an example. As shown in Fig. \ref{fig.6}(a), we analyze the interference fringe with the radius $r=1.7$ mm. According to Eq. \ref{Eq.(5)}, we use a cosine function $I=I_0+A\times\cos[-\Delta l(\phi-\phi_{max})]$ to fit the data. Here $\Delta l= 2 l_{1}=-6$. $\phi_{max}$ is the azimuthal angle for the maximum intensity of the first bright interference fringe, and its value is set in the range $[-\frac{\pi}{|\Delta l|}, \frac{\pi}{|\Delta l|}]$. It is noted that to optimize the fitting, two circles with $\phi \in (0,\ 4\pi)$ are plotted. From the fitting, $\phi_{max}=0.086$ rad. Then we plot $\phi_{max}$ as a function of $r$ in Fig. \ref{fig.6}(b). According to Eq. \ref{Eq.(9)}, we use a quadratic function $\phi_{max}(r)=\alpha r^2+\phi_c$ to fit the data, extracting $\alpha=8.05(17)\times10^4$ rad m$^{-2}$. The data in the parenthesis is the standard deviation from the fitting. From the calculation of Eq. \ref{Eq.(10)}, $\alpha=8.30\times10^4$ rad m$^{-2}$. The experimental measurement agrees well with the theoretical calculation.

In Fig. \ref{fig.7}, we measure $\alpha$ for different TCs. For $l_{1} < 0$, $\alpha$ decreases with increasing $\Delta z$. If the magnitude of $l_{1}$ is big, $\alpha$ varies slowly. If there is no wavefront curvature difference ($\Delta z = 0$), $\alpha$ is vanishing small. The sign of $\alpha$ changes when $\Delta z$ varies from the negative to positive, which is similar to the observations in Fig. \ref{fig.5}. The measurements are consistent to the theoretical calculation of Eq. \ref{Eq.(10)}.

Several previous works have observed the twist effects of the interference pattern mainly based on the visual inspection \cite{Harris1994OCspiral, Cui2019OE, Vickers2008OSA}. Here we quantitatively extract the value of the parameter $\phi_{max}$ from the interference fringe and plot the variation of $\phi_{max}$ along the radial direction. We further measure the twist coefficient $\alpha$ to characterize the degree of the radial twist. Effects of the wavefront curvature difference (denoted with the distance difference $\Delta z$) as well as the TC $l_1$ on the radial twist are systematically explored.

\section{Conclusion}
In conclusion, compared to the previous works mainly on the visual inspection, we quantitatively characterize the temporal rotation and radial twist of the interference pattern of vortex beam. The effects of the physical parameters (i.e., the TC, frequency difference and wavefront curvature difference of the vortex beams) on the temporal rotation as well as the radial twist are systematically explored. We measure two parameters, the rotation velocity $\Omega$ and twist coefficient $\alpha$, respectively, to characterize the degree of the temporal rotation and radial twist. The method demonstrated here is also applicable to other kinds of vortex beams. The knowledge of the detailed structure of the interference pattern can extend the application in manipulation of matter using superimposed vortex beams \cite{Hu2021PhysicsVortices, Aldossary2017JOSAArtificial, Peter2016PRAHall, Zhan2009OEZhan}.

\section*{Acknowledgments}

This work has been supported by the NKRDP (National Key Research and Development Program) under Grant No. 2016YFA0301503, NSFC (Grant No. 11674358, 11904388, 12004398, 12121004), CAS under Grant No. YJKYYQ20170025 and K. C. Wong Education Foundation (Grant No. GJTD-2019-15), and Hubei province under Grant No. 2021CFA027. Longzhi Nie and Lingran Kong contributed equally to this work.

\section*{Declaration of competing interest}

The authors declare that they have no known competing financial interests or personal relationships that could have appeared to influence the work reported in this paper.


\section*{References}

\begin{thebibliography}{10}
\expandafter\ifx\csname url\endcsname\relax
  \def\url#1{\texttt{#1}}\fi
\expandafter\ifx\csname urlprefix\endcsname\relax\def\urlprefix{URL }\fi
\expandafter\ifx\csname href\endcsname\relax
  \def\href#1#2{#2} \def\path#1{#1}\fi

\bibitem{Allen1992PRAoam}
L.~Allen, M.~W. Beijersbergen, R.~J.~C. Spreeuw, J.~P. Woerdman, Orbital
  angular momentum of light and the transformation of {Laguerre-Gaussian} laser
  modes, Phys. Rev. A 45 (1992) 8185--8189.

\bibitem{Padgett1995OC}
M.~J. Padgett, L.~Allen, The poynting vector in {Laguerre-Gaussian} laser
  modes, Opt. Commun 121~(13) (1995) 36--40.

\bibitem{Bozinovic2013Science}
N.~Bozinovic, Y.~Yue, Y.~Ren, P.~K. M.~Tur, H.~Huang, A.~E. Willner,
  S.~Ramachandran, Terabitscale orbital angular momentum mode division
  multiplexing in fibers, Science 340~(6140) (2013) 1545--1548.

\bibitem{Vallone2014prl}
G.~Vallone, V.~D. Ambrosio, A.~Sponselli, S.~Slussarenko, L.~Marrucci,
  F.~Sciarrino, P.~Villoresi, Freespace quantum key distribution by
  rotation-invariant twisted photons, Phys. Rev. Lett 113~(6) (2014) 060503.

\bibitem{Yao2011Adv}
A.~M. Yao, M.~J. Padgett, Orbital angular momentum origins, behavior and
  applicationss, Adv. Opt. Photonics 3~(2) (2011) 161--204.

\bibitem{Grier2003Nature}
D.~G. Grier, A revolution in optical manipulation, Nature 424~(6950) (2003)
  810--816.

\bibitem{Paterson2001Science}
L.~Paterson, M.~P. MacDonald, J.~Arlt, W.~Sibbett, P.~E. Bryant, K.~Dholakia,
  Controlled rotation of optically trapped microscopic particles, Science
  292~(5518) (2001) 912--914.

\bibitem{Lehmuskero2014OE}
A.~Lehmuskero, Y.~Li, P.~Johansson, M.~K\"{a}ll, Plasmonic particles set into
  fast orbital motion by an optical vortex beam, Opt. Express 22~(4) (2014)
  4349--4356.

\bibitem{Pu2015PRAsoam}
M.~DeMarco, H.~Pu, Angular spin-orbit coupling in cold atoms, Phys. Rev. A 91
  (2015) 033630.

\bibitem{Zhang2015PRAsoam}
K.~Sun, C.~Qu, C.~Zhang, Spin--orbital-angular-momentum coupling in
  {Bose-Einstein} condensates, Phys. Rev. A 91 (2015) 063627.

\bibitem{Lin2018PRLsoam}
H.-R. Chen, K.-Y. Lin, P.-K. Chen, N.-C. Chiu, J.-B. Wang, C.-A. Chen, P.-P.
  Huang, S.-K. Yip, Y.~Kawaguchi, Y.-J. Lin, Spin-orbital-angular-momentum
  coupled {Bose-Einstein} condensates, Phys. Rev. Lett. 121 (2018) 113204.

\bibitem{Lin2018PRLrotation}
P.-K. Chen, L.-R. Liu, M.-J. Tsai, N.-C. Chiu, Y.~Kawaguchi, S.-K. Yip, M.-S.
  Chang, Y.-J. Lin, Rotating atomic quantum gases with light-induced azimuthal
  gauge potentials and the observation of the {Hess-Fairbank} effect, Phys.
  Rev. Lett. 121 (2018) 250401.

\bibitem{Jiang2019PRLSOAMC}
D.~Zhang, T.~Gao, P.~Zou, L.~Kong, R.~Li, X.~Shen, X.~Chen, S.~Peng, M.~Zhan,
  H.~Pu, K.~Jiang, Ground-state phase diagram of a
  spin-orbital-angular-momentum coupled {Bose-Einstein} condensate, Phys. Rev.
  Lett. 122 (2019) 110402.

\bibitem{Leach2007OE}
S.~Franke-Arnold, J.~Leach, M.~J. Padgett, V.~E. Lembessis, D.~Ellinas, A.~J.
  Wright, J.~M. Girkin, P.~\"{O}hberg, A.~S. Arnold, Optical ferris wheel for
  ultracold atoms, Opt. Express 15~(14) (2007) 8619--8625.

\bibitem{Arlt2002OEMovingDoppler}
J.~Arlt, M.~MacDonald, L.~Paterson, W.~Sibbett, K.~Dholakia,
  K.~Volke-Sepulveda, Moving interference patterns created using the angular
  doppler-effect, Opt. Express 10~(16) (2002) 844--852.

\bibitem{Harris1994OCspiral}
M.~Harris, C.~Hill, J.~Vaughan, Optical helices and spiral interference
  fringes, Opt. Commun. 106~(4) (1994) 161--166.

\bibitem{Hu2021PhysicsVortices}
G.~Liang, B.~Yuan, Y.~Li, X.~Kong, W.~Cheng, H.~Qiao, X.~Hu, Evolutions of
  optical vortices under wide gaussian background, Results Phys. 26 (2021)
  104352.

\bibitem{Liu2021OpticsRadius}
D.~Yang, Z.~Yang, Z.~Zhao, Z.~Liu, Radius of curvature of spherical wave
  measurement based on vortex beam interference, Opt. Lasers Eng. 142 (2021)
  106592.

\bibitem{Aldossary2017JOSAArtificial}
V.~E. Lembessis, A.~Alqarni, S.~Alshamari, A.~Siddig, O.~M. Aldossary,
  Artificial gauge magnetic and electric fields for free two-level atoms
  interacting with optical {Ferris} wheel light fields, J. Opt. Soc. Am. B
  34~(6) (2017) 1122--1129.

\bibitem{Peter2016PRAHall}
M.~Lacki, H.~Pichler, A.~Sterdyniak, A.~Lyras, V.~E. Lembessis, O.~Al-Dossary,
  J.~C. Budich, P.~Zoller, Quantum hall physics with cold atoms in cylindrical
  optical lattices, Phys. Rev. A 93 (2016) 013604.

\bibitem{Zhan2009OEZhan}
X.~He, P.~Xu, J.~Wang, M.~Zhan, Rotating single atoms in a ring lattice
  generated by a spatial light modulator, Opt. Express 17~(23) (2009)
  21007--21014.

\bibitem{Cui2019OE}
S.~Cui, B.~Xu, S.~Luo, H.~Xu, Z.~Cai, Z.~Luo, J.~Pu, S.~Ch\'{a}vez-Cerda,
  Determining topological charge based on an improved {Fizeau} interferometer,
  Opt. Express 27~(9) (2019) 12774--12779.

\bibitem{Vickers2008OSA}
J.~Vickers, M.~Burch, R.~Vyas, S.~Singh, Phase and interference properties of
  optical vortex beams, J. Opt. Soc. Am. A 25~(3) (2008) 823--827.

\bibitem{Khajavi2017OL}
B.~Khajavi, E.~J. Galvez, Determining topological charge of an optical beam
  using a wedged optical flat, Opt. Lett 42~(8) (2017) 1516--1519.

\bibitem{Riley1977ApplOpt}
M.~E. Riley, G.~M. A, Laser beam divergence utilizing a lateral shearing
  interferometer., Appl. Opt. 16~(10) (1977) 2753--2756.

\bibitem{Chappuis2017pra}
R.~Geneaux, C.~Chappuis, T.~Auguste, Radial index of {Laguerre-Gaussian} modes
  in high-order-harmonic generation, Phys. Rev. A 95~(5) (2017) 051801.

\bibitem{Lembessis2017pra}
V.~E. Lembessis, E.~Vasileios, Atomic ferris wheel beams, Phys. Rev. A 96~(1)
  (2017) 013622.

\bibitem{Baumann2009OE}
S.~M. Baumann, D.~M. Kalb, L.~H. MacMillan, E.~J. Galvez, Propagation dynamics
  of optical vortices due to {Gouy} phase, Opt. Express 17~(12) (2009)
  9818--9827.

\bibitem{Nie2015OCmeasureTC}
X.~Li, Y.~Tai, F.~Lv, Z.~Nie, Measuring the fractional topological charge of
  {LG} beams by using interference intensity analysis, Opt. Commun. 334 (2015)
  235--239.

\end{thebibliography}
\end{document}